\documentclass[a4paper,aps,twocolumn,nobibnotes,pra]{revtex4-1}
\usepackage[T1]{fontenc}
\usepackage[utf8]{inputenc}
\usepackage{lmodern}
\usepackage{graphicx}
\usepackage{dcolumn}
\usepackage{bm}
\usepackage{amsmath}
\usepackage{amstext}
\usepackage{amssymb}
\usepackage{amsthm}
\usepackage{dsfont}
\usepackage{color}
\usepackage[caption=false]{subfig}
\usepackage{tabularx}
\usepackage{booktabs}
\usepackage[table]{xcolor}
\usepackage{colortbl}
\definecolor{lightgray}{rgb}{0.85,0.85,0.85}

\usepackage[bookmarks=false, hidelinks]{hyperref}
\hypersetup{colorlinks=true,citecolor=blue,linkcolor=blue,%
urlcolor=blue,pdfstartview=FitH,bookmarksopen=true}

\newcommand{\ket}[1]{\ensuremath{| #1 \rangle}}

\newcommand{\ketbra}[1]{\ensuremath{| #1 \rangle\langle #1 |}}

\newcommand{\id}[0]{\ensuremath{\mathds{1}}}

\newcommand{\mm}[1]{\begin{align} #1 \end{align}}

\newcommand{\be}{\begin{equation}}
\newcommand{\ee}{\end{equation}}

\newcommand{\bea}{\begin{eqnarray}}
\newcommand{\eea}{\end{eqnarray}}

\renewcommand{\rho}{\varrho}
\renewcommand{\tilde}{\widetilde}

\theoremstyle{definition}

\newcommand{\expect}[1]{\langle #1 \rangle}

\newcommand{\conv}{\ensuremath \text{conv}}

\newcommand{\R}{\mathbb R}

\newcommand{\refeq}[1]{Eq.~(\ref{#1})}

\begin{document}

\title{Bell inequalities for nonlocality depth}

\author{Fabian Bernards}
\author{Otfried Gühne}%
\affiliation{
Naturwissenschaftlich-Technische Fakultät, 
Universität Siegen, 
Walter-Flex-Straße 3, 
57068 Siegen, Germany}

\date{\today}

\begin{abstract}
When three or more particles are considered, quantum correlations 
can be stronger than the correlations generated by so-called hybrid 
local hidden variable models, where some of the particles are 
considered as a single block inside which communication and signaling
is allowed. We provide an exhaustive classification of Bell inequalities 
to characterize various hybrid scenarios in four- and five-particle systems. 
In quantum mechanics, these inequalities provide device-independent witnesses 
for the entanglement depth. In addition, we construct a family of inequalities 
to detect a non-locality depth of $(n-1)$ in $n$-particle systems. Moreover, 
we present two generalizations of the original Svetlichny inequality, which was 
the first Bell inequality designed for hybrid models. Our results are based on 
the cone-projection technique, which can be used to completely characterize
Bell inequalities under affine constraints; even for many parties, measurements, 
and outcomes. 
\end{abstract}

\maketitle

\section{Introduction}
Bell inequalities have been successfully used to show that the results 
of Bell test experiments are incompatible with  local hidden-variable 
(LHV) models \cite{brunner_bell_2014, shalm_strong_2015,
giustina_significant-loophole-free_2015,
hensen_loophole-free_2015, rosenfeld_event-ready_2017}. 
Moreover, nonlocality has been linked to communication complexity problems 
\cite{buhrman_nonlocality_2010} and Bell inequalities have been found useful 
for device independent verification of quantum states and measurements 
\cite{supic_self-testing_2020}.

\begin{figure}[t!]
    \centering
    \subfloat[]{\includegraphics[width=.20\columnwidth]{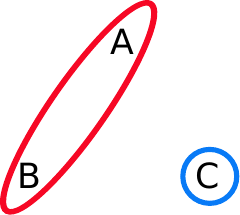}}
    \hspace{.1\columnwidth}
    \subfloat[]{\includegraphics[width=.20\columnwidth]{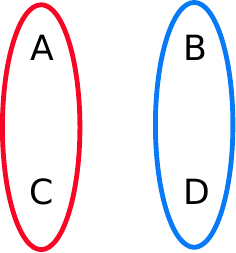}}
    \hspace{.1\columnwidth}
    \subfloat[]{\includegraphics[width=.20\columnwidth]{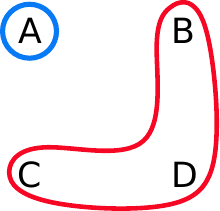}}
    \caption{Illustration of different hybrid models. (a) In the scenario
    considered by Svetlichny, in each round of the Bell experiment two of the
    three parties (here, Alice and Bob) form a team and are allowed to communicate. 
    In (b) and (c) we see two different hybrid models for four parties. In each round 
    of the experiment, there could be two teams of two parties each (b) or one team of 
    three parties and an isolated party (c). The behaviors that can arise from scenario 
    (b) form a so-called $(2,2)$ model, scenario (c) gives rise to a $(3,1)$ model.}
    \label{fig-models}
\end{figure}

Given the phenomenon of quantum nonlocality, one may ask whether for 
three or more particles novel effects occur. To answer this question,
George Svetlichny introduced in 1987 so-called hybrid models
\cite{svetlichny_distinguishing_1987}. Hybrid models are a class 
of hidden variable models that give up on all restrictions on the
correlations between a subset of parties while maintaining the 
restriction of local realism with respect to the remaining ones. 
The most famous example of a hybrid model is the one originally 
introduced by Svetlichny for three parties: In any round of the Bell 
experiment, two of the parties may collaborate to establish arbitrary 
correlations between themselves (see also Fig.~\ref{fig-models}). However, the correlations shared between 
these two parties and the third party must respect a LHV model. In this 
case, the correlations between the three parties satisfy the Svetlichny 
inequality \cite{svetlichny_distinguishing_1987}, which reads
\mm{
    \expect{A_1 B_1 C_2} + \expect{A_1 B_2 C_1} + \expect{A_2 B_1 C_1}- \expect{A_2
 B_2 C_2}&\notag\\ + \expect{A_2 B_2 C_1} + \expect{A_2 B_1 C_2} + \expect{A_2
 B_1 C_1} - \expect{A_1 B_1 C_1} &\le 4,
 }
where $A_a, B_b, C_c$ denote measurements on parties A, B, and C, respectively,
and each of the measurements yields outcomes $\pm 1$. In quantum mechanics, 
the Svetlichny inequality is maximally violated up to a value of $4 \sqrt 2$ by 
the Greenberger-Horne-Zeilinger (GHZ) state
\mm{
    \ket{\text{GHZ}} = \frac 1{\sqrt 2} (\ket{000} + \ket{111} ).
}
the Svetlichny inequality has been generalized to more particles
\cite{collins_bell_2002}, and experimental violations of these inequalities have also been 
observed \cite{lavoie_experimental_2009, zhao_experimental_2003}.

For our further discussion it will be convenient to have a compact notation.
By defining $[abc] := \langle A_a B_b C_c \rangle$, we can write the Svetlichny
inequality as
\mm{
4 &- [112] - [121] - [211] + [222] \notag\\
  &- [221] - [212] - [122] + [111] \ge 0.
}
We have mentioned before that a violation of the Svetlichny inequality implies
genuine tripartity nonlocality that cannot be explained by a hybrid model. 
Let us shortly explain why this is the case. Assume that Alice and 
Bob can share arbitrary correlations,  since the inequality is symmetric 
under exchange of parties this is no restriction of generality. Allowing 
arbitrary correlations among two parties amounts to treating both together 
as one party. Rewriting the Svetlichny inequality in this way yields
\mm{
4 - [12] - [21] - [31] + [42] - [41] - [32] - [22] + [11] \ge 0.
}
This inequality, however, can be written as the sum of two Clauser-Horne-Shimony-Holt 
(CHSH) inequalities \cite{clauser_proposed_1969, clauser_proposed_1970},
\mm{
2 - [12] - [21] - [22] + [11] \ge &0 \label{eq-svetchsh1},\\
2 - [31] + [42] - [41] - [32] \ge &0\label{eq-svetchsh2}.
}
Consequently, for models which are local with respect to the $AB|C$
partition, the Svetlichny inequality holds, and violation of it requires
at least one of the two CHSH inequalities \refeq{eq-svetchsh1}, \refeq{eq-svetchsh2} to 
be violated. Due to the permutation symmetry of the Svetlichny inequality this 
argument shows that a violation of the Svetlichny inequality implies that any 
two parties share nonlocality with the remaining one.

As mentioned earlier, the Svetlichny inequality is violated in quantum mechanics, 
which means that nonlocality affects more than two particles at once. To gain 
a better understanding of this phenomenon, it is desirable to find other inequalities
for hybrid scenarios. In the scenario considered by Svetlichny, there is only one 
full-body correlation hybrid inequality, which is different from the Svetlichny 
inequality, found by Jean Daniel Bancal et al.~\cite{bancal_looking_2010}. Moreover, 
hybrid Bell inequalities have been derived for an arbitrary number of parties 
\cite{collins_bell_2002}, an arbitrary number of parties with two dichotomic 
measurements \cite{bancal_detecting_2011} and also scenarios with $n$ parties 
that have access to $m$ measurements with $o$ outcomes each 
\cite{bancal_framework_2012}. Beyond this, scenarios have been considered 
in which some parties are grouped together, but within such as group still all
correlations in the hidden-variable models obey the nonsignaling constraint
 \cite{almeida_multipartite_2010}.

The question arises, how one can find Bell inequalities for hybrid models 
of four and five parties. For more than three parties there is more
than one hybrid model to consider (see Fig.~\ref{fig-models}).
For four parties, one example of a hybrid model would be one, where two 
teams of two parties share local correlations while the correlations shared 
between parties within a team can be arbitrary. In the following we refer to 
this hybrid model as a $(2,2)$ model. Another example would be a hybrid model 
where there is one team of three parties and one party that is on its own. In 
our terminology, this is a $(3,1)$ model.

In this paper, we consider all hybrid models for four and five parties and find 
{\it all} optimal, symmetric, full-body correlation Bell inequalities, for two 
settings with two outcomes per observer. Full-body correlation Bell inequalities 
are those that only involve correlations where every party performs a non-trivial 
measurement. Such inequalities have been considered in detail for the usual notion
of fully local hidden variable models \cite{zukowski_bells_2002,
werner_all_2001}.
Requiring symmetry of the Bell inequalities imposes linear contraints, which can 
be incorporated using the cone-projection technique (CPT)
\cite{bernards_generalizing_2020, bernards_finding_2021}. 
This technique helps to reduce the dimensionality of the problem and thus simplifies 
the task and allows us to find {\it all} Bell inequalities. Finally, we then shift 
our focus back to three-partite nonlocality, where we discuss genuine-multipartite 
Bell inequalities with three settings that generalize the Svetlichny inequality. 
This means that for a particular choices of measurements, these inequalities 
reduce to the Svetlichny inequality. Again, the CPT allows to tackle this problem.

\section{Hybrid models}

Consider a Bell experiment where the parties perform local measurements
that yield outcomes $\pm 1$. The parties can then describe the behavior 
of the experiment by estimating the expectation values for all combinations 
of measurements, such as $\langle A_a B_b C_c \rangle$. Such an expectation 
value is called a {\it correlation}. A {\it behavior} is a vector that encodes 
the information of some or all of the correlations measured in the experiment. 
Let $d$ be the number of correlations that are measured in the experiment. 
Then, the euclidian space $\R^d$ can be used to describe all the information 
encoded in the behaviours. 
A  physical model, such as quantum mechanics, an LHV model or a hybrid 
model is a subset of the vector space of behaviors. Moreover, every model
is a subset of the hypercube defined by the conditions
\mm{
-1 \le \langle A_a B_b C_c \rangle \le 1.
}
We refer to this hypercube as the unconstrained model $M_{\square}$.
Hybrid models are models that are more restrictive than the 
unconstrained model but less restrictive than LHV models 
\cite{svetlichny_distinguishing_1987, collins_bell-type_2002}. 
For the set of hybrid models $M_H$  it thus holds that 
\mm{
M_{LHV} \subset M_H \subset M_{\square}, 
}
where $M_{LHV}$ is the set of LHV model and $M_{\square}$ is the set
of unconstrained behaviors.

Hybrid models can be constructed in the following way: Given an $n$-party 
system with subsystems $s \in S$, the first step is to define a partition 
\mm{
P &= \{c_1, \ldots, c_k\}
}
where the cells $c_i \subset S$ are disjoint and non-empty subsets 
of $S$ and $\bigcup_{i = 1}^k c_i = S$.

In a second step, one defines a LHV model $M_{LHV}^P$ for the 
coarse grained scenario defined by $P$. This works in the following way: 
Every cell $c_i$ of the partition is considered as one system. The measurement 
settings are all combinations of measurement settings that apply to each 
subsystem within a cell. However, no restrictions apply to the correlations 
between subsystems within a cell, since the cell is regarded as one system. In
particular, this allows for signaling to take place between the parties within
one cell.

As a third step, two partitions are considered equivalent, if they are related
by a relabeling of the parties. Each equivalence class of partitions of a
partition $P$ is then defined by what we call the cardinality tuple
\mm{
h_P = (|c_i| \mid i \in\{1, \ldots, k\}, \, c_i \in P ),
}
which contains the ordered cardinalities of the cells of $P$. Two partitions 
$P, P'$ are equivalent if and only if $h_P = h_{P'}$. 
For any ordered tuple $h$, one defines a hybrid model  
\mm{
M_h = \conv \Big(\bigcup_{P \mid h_P = h} M_{LHV}^P\Big),
}
where $\conv$ denotes the convex hull. Note that $M_h$ is a convex polytope, 
the extremal points of which are the union of the extremal points of models
$M_{LHV}^P$.

If $h$ is an $m$-tuple, we call $M_h$ an $m$-local model and if $m$ is equal 
to the number of parties, we call the resulting model \textit{fully local}.
A more detailed discussion on different notions of multipartite nonlocality can
be found in Refs.~\cite{szalay_k-stretchability_2019, baccari_bell_2019}.
Some authors have for example considered hybrid models that impose
a no-signaling constraint on the behaviors of each cell in a given partition
\cite{almeida_multipartite_2010, bancal_definitions_2013}. 
In this setup, device independent certification of entanglement has been
investigated \cite{liang_family_2015, aloy_device-independent_2019,
lin_exploring_2019, tura_optimization_2019}. For four parties and $h_P = (2,2)$
all Bell-inequalities that are symmetric under party permutations have been
found \cite{curchod_multipartite_2014}. Moreover, it is known, that for full-body
correlation Bell inequalities, the additional no-signaling constraint on the
parties within each cell does not make a difference
\cite{curchod_quantifying_2015}.

Note that there are also other approaches to nonlocality, including the
so-called operational
 approach \cite{gallego_operational_2012}. Recently, genuine multipartite nonlocality and
entanglement have
also been studied in networks. This leads to definitions that are
different from genuine multipartite nonlocality as introduced by Svetlichny
\cite{navascues_genuine_2020, hansenne_symmetries_2022, tavakoli_bell_2022}.

\section{Hybrid Bell inequalities}

\subsection{Statement of the problem}

We consider the case of four and five parties that seek to perform a Bell
experiment in order to investigate the structure of the nonlocality they might
share. Every party can choose between two measurement settings, each of which
yields outcomes $\pm 1$. More specifically, we consider the case in which every 
party performs one of the two measurements in every round, that is, every party
performs a non-trivial measurement in every round. As far as Bell inequalities 
are concerned, this means that we only consider full-body correlation Bell 
inequalities. It is worth mentioning that for fully local models, all full-body 
correlation inequalities are known \cite{werner_all_2001, zukowski_bells_2002}. 
Further, we only consider Bell inequalities that are symmetric under relabeling 
of the parties. This symmetry constraint is a linear constraint on the coefficients 
of the Bell inequality in question, so we can employ the cone-projection technique 
\cite{bernards_finding_2021} to specifically find these Bell inequalities.

In order to be able to investigate the nonlocal structure, we need to consider
different hybrid models. In the four-party case, these hybrid models are given 
by the cardinality tuples
\mm{ h \in \{(1,1,1,1), (2,1,1), (2,2), (3,1)\},
}
where $h = (1,1,1,1)$ corresponds to the fully local model. In the case of five 
parties, we consider six models given by the cardinality tuples 
\mm{
h \in \{&(1,1,1,1,1), (2,1,1,1), (2,2,1), \\
&(3,1,1),(3,2), (4,1)\}.}
For every
model, we find all optimal, symmetric, full-body correlation Bell inequalities.

To aid readability, we introduce some notation that simplifies writing down Bell
inequalities which are symmetric under permutation of the parties. For example,
for four parties $A, B, C, D$ we write
\mm{
(1122) &= \expect{A_1 B_1 C_2 D_2} + \text{party permutations}
\label{eq-symmnotation},
\intertext{where 'party permutations' only includes permutations that yield
different terms. Therefore, expression \refeq{eq-symmnotation} consists of six
terms. To give another example,
}
(1112) &= \expect{A_1 B_1 C_1 D_2} + \expect{A_1 B_1 C_2 D_1}\notag\\
&\quad +\expect{A_1 B_2
C_1 D_1} +\expect{A_2 B_1 C_1 D_1}.
}
For five parties, the notation works analogously.

\subsection{Description of the method}

We find the Bell inequalities using the cone-projection technique (CPT), which 
is introduced and and in detail described in Refs.~\cite{bernards_generalizing_2020, 
bernards_finding_2021}. The CPT is a method to completely characterize facet-defining 
Bell inequalities that obey a set of affine equality constraints. Naively, one may 
achieve this by simply enumerating all facet-defining Bell inequalities and select 
those which meet the criteria in a second step. However, the CPT provides is a more 
elegant and efficient to achieve this goal that even works in cases in which finding 
all Bell inequalities is infeasible. 

The CPT consists of three steps. In the first step, 
on disregards the normalization and associates a ray with every extremal behavior (that is, 
a point in the space of all behaviours). Conversely, the extremal behaviors can be recovered 
from the rays by intersecting with a hyperplane. These rays generate a cone $C$. In the second
step, the constraints on the Bell inequalities define a lower-dimensional subspace, into 
which $C$ is projected. This yields a cone $\tilde C$. In the third step, one finds
all Bell inequalities that meet the constraints by enumerating the facet-defining 
inequalities of $\tilde C$ and checking which of these indeed correspond to facets of $C$. 
These inequalities are exactly the facet defining Bell inequalities that obey the
constraints.

\subsection{Numerical analysis of the Bell inequalities}

For each Bell inequality, we perform the same numerical analysis. We 
find a lower bound on the quantum violation using qubit and qutrit 
systems and an upper bound using the third level of the NPA-hierarchy 
\cite{navascues_bounding_2007, navascues_convergent_2008}. We also find 
the no-signaling bound. This is a linear program, which can be solved 
using, for example, the MOSEK solver \cite{noauthor_mosek_nodate}. We find that while 
not all inequalities are violated in quantum mechanics, all can be 
violated using no-signaling behaviors. We calculate the quantum violation 
of each Bell inequality using a seesaw algorithm that optimizes the
settings of one party in every step and cycles through the parties. This
algorithm is not guaranteed to yield the maximal quantum violation. However,
comparing with the upper bound provided by the NPA-hierarchy, we can 
confirm that the optimum was achieved (within numerical accuracy) in all 
cases.  All of bounds for the inequalities are listed together with the 
inequalities in the Supplementary Material, see also the description in Appendix~\ref{app-supp}.

Additionally, we find lower bounds on the noise robustnesses of the 
inequalities using the states
\mm{
\rho_4(p) &= (1-p) |\text{GHZ}_4\rangle \langle GHZ_4| + \frac p{16} \id \label{eq-rho4}
\intertext{and for five qubits}
\rho_5(p) &= (1-p) |\text{GHZ}_5\rangle \langle GHZ_5| + \frac p{32} \id, \label{eq-rho5} 
\intertext{where the respective GHZ states are given by}
|\text{GHZ}_4\rangle &= \frac 1{\sqrt 2} ( |0000\rangle + |1111\rangle ), \\
|\text{GHZ}_5\rangle &= \frac 1{\sqrt 2} ( |00000\rangle + |11111\rangle ).
}
We identify regimes of the parameter $p$, for which some of the
hybrid models can be excluded, details are described in 
Appendix~\ref{app-interv}.

\subsection{Four-party nonlocality}

\begin{figure*}[t!]
\includegraphics[width=\textwidth]{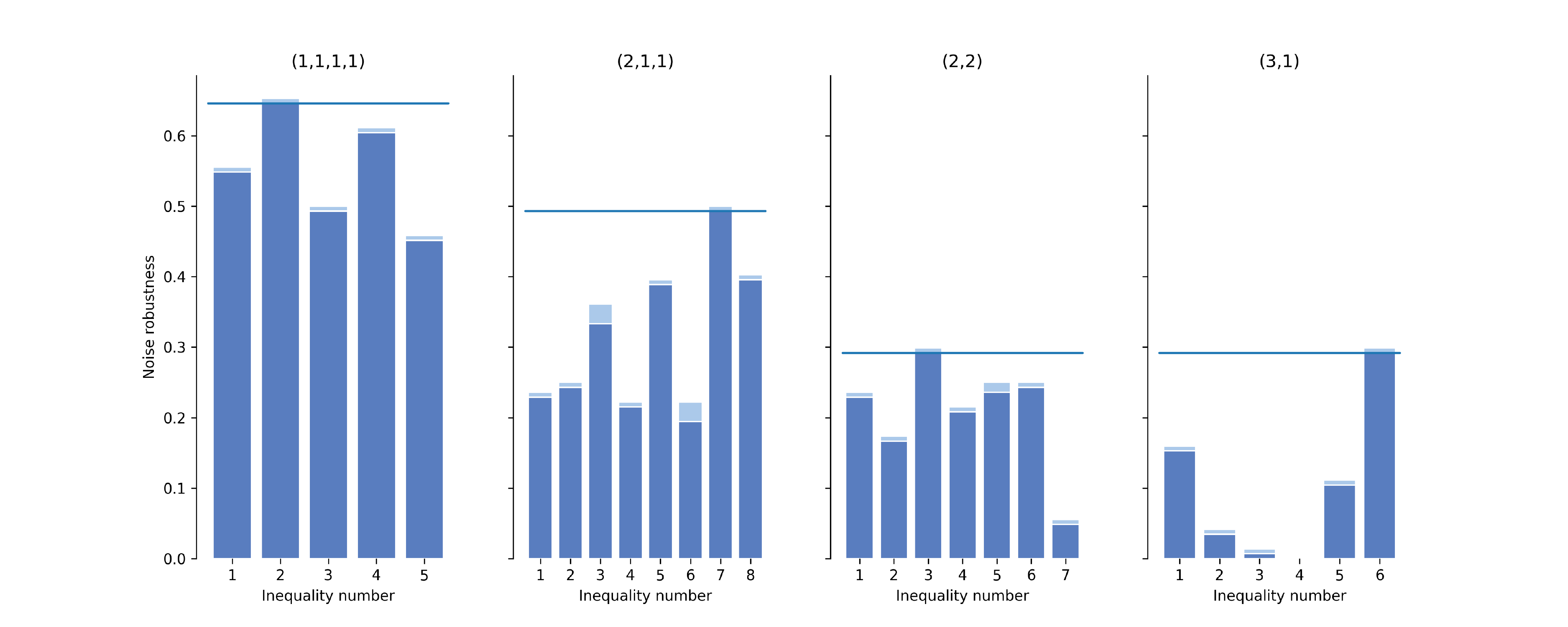}
\caption{Noise robustness for each inequality for hybrid models in the four-party case, labeled
by the respective model. Details can be found in the Supplementary Material A, B, C, and D. Given the family of states 
$\rho_4(p)$ as defined in \refeq{eq-rho4} the dark-blue (dark-gray) 
bar indicates a value of $p$ for which the violation is larger than some threshold (here: $10^{-6}$), 
the light-blue  (light-gray) bar indicates a value of $p$ for which the threshold is no longer exceeded. 
Note that inequality 4 in the $(3,1)$ model is not violated in quantum mechanics at all.
In each plot, the horizontal bar indicates the highest achieved noise robustness.
}
\label{fig-nr1}
\end{figure*}

\begin{table*}[t]
\begin{tabular}{l|l|l|l|l}
\toprule
      model & \#ineq. & noise-robustness & ineq. number &
inequality \\
\midrule
 (1,1,1,1)  &            5 &         0.645 &                 2 &   
$+4 - (1111) - (1112) + (1122) + (1222) - (2222)$ \\
   (2,1,1)  &            8 &         0.493 &                 7 &
$+4 - (1112) + (1222)$ \\
     (2,2)  &            7 &         0.291 &                 3 &
$+16 - (1112) -2 (1122) +3 (1222) +4 (2222)$ \\
     (3,1)  &            6 &         0.291 &                 6 &  
$ +8 + (1111) - (1112) - (1122) + (1222) + (2222)$ \\
\bottomrule
\end{tabular}
\label{tab:fourparty}
\caption{For each hybrid model, the table shows the number of optimal symmetric
full-body correlation inequalities. Further a lower bound for the white-noise
robustness for the $|GHZ\rangle$ state is provided for the best inequality,
which is labeled by its number in the list in the Supplementary Material.}
\end{table*}

In the four-party case, we find in total 26 Bell inequalities, five for the
fully local model, eight for the $(2,1,1)$ model, seven for the $(2,2)$ model,
and six for the $(3,1)$ model. Of these inequalities, we find all but one to be
violated in quantum mechanics. All inequalities that are violated in quantum
mechanics, are maximally violated by the GHZ state.

For the fully local model, the inequality that exhibits the best white-noise
robustness with respect to the GHZ-state is the generalized Mermin inequality
\mm{
+4 - (1111) - (1112) + (1122) + (1222) - (2222) \ge 0,
}
which was already found in Ref.~\cite{collins_bell-type_2002}. With it, 
nonlocality can be detected with up to roughly $64.5\%$ white noise.

One might expect that for the hybrid models considered, one would find that
the Bell inequality with the best noise robustness is a generalized Svetlichny
inequality. However, this is not the case.
For the $(2,1,1)$ model, the most noise robust Bell inequality is 
\mm{
4 - (1112) + (1222) \ge 0
}
and it is violated up to roughly $49.3\%$ of white noise.

If the amount of white noise is less than roughly $p = 29.1\%$, then the state
$\rho_4(p)$ violates the
inequality
\mm{
+8 + (1111) - (1112) - (1122) + (1222) + (2222) \ge 0
}
for the $(3,1)$ model and the inequality
\mm{
+16 - (1112) -2 (1122) +3 (1222) +4 (2222) \ge 0  \label{eq-newine4}
}
for the $(2,2)$ model. Interestingly, the amount of white-noise required to
obtain a violation of the $(2,2)$ model and the $(3,1)$ model is the same for
symmetric, full-body correlation, two-setting inequalities, although the
violations are established by different inequalities. We will observe the same
phenomenon in the case of five parties. The findings discussed in this
subsection are summarized in Table \ref{tab:fourparty}. The noise robustnesses of
all inequalities we found in the four-party case are plotted in Figure
\ref{fig-nr1}.

\subsection{Five-party nonlocality}

\begin{figure*}[p!]
\includegraphics[width=\textwidth]{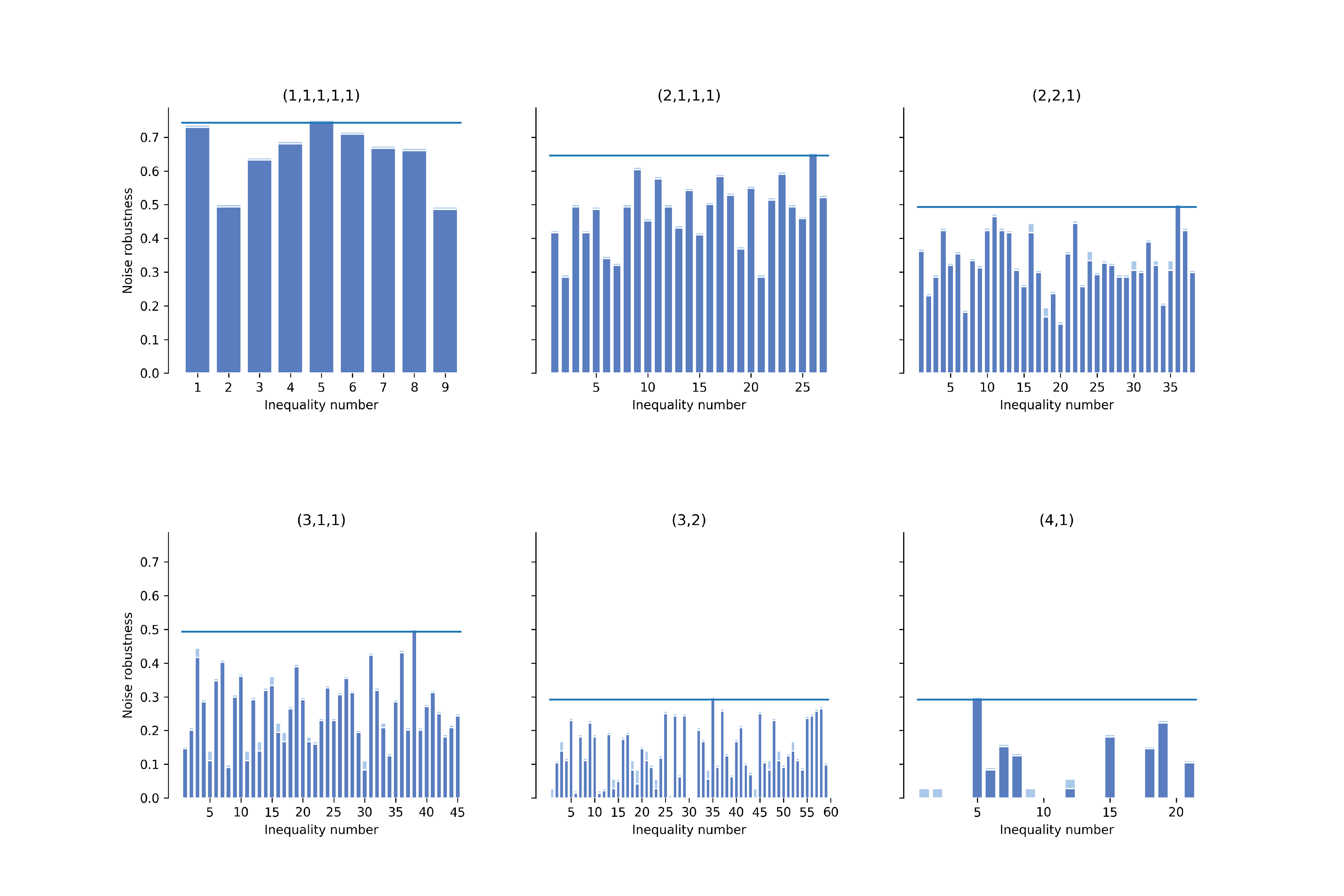}
\caption{Noise robustness for each inequality for hybrid models in the five-party case, labeled 
by the respective model. Details can be found in the Supplementary Material E, F, G, H, I, and J. The dark-blue (dark-gray) bar indicates a value
of $p$ for which the violation is larger than some threshhold (here: $10^{-6}$),
the light-blue (light-gray) bar indicates a value of $p$ for which the
threshhold is no longer exceeded. Note that inequalities 30 and 31 of the 3-2
model as well as inequalities 3, 4, 10, 11, 13, 14, 16, 17 and 20 for the 4-1
local model are not violated in quantum mechanics. }
\label{fig-nr2}
\end{figure*}
\begin{table*}[p!]
\begin{tabular}{l|l|l|l|l}
\toprule
        model & \#ineq. & noise-robustness & ineq. number &
inequality \\
\midrule
 (1,1,1,1,1)  &            9 &         0.743 &                 5 &
$+4 - (11112) + (11222) - (22222)$ \\
   (2,1,1,1)  &           27 &         0.645 &                26 &    +8
$+ (11111) + (11112) - (11122) - (11222) + (12222) + (22222)$ \\
    (2,2,1) &           38 &         0.493 &                36 &
$+8 + (11111) - (11122) + (12222)$ \\
    (3,1,1) &           45 &         0.493 &                38 &
$+8 + (11111) - (11122) + (12222)$ \\

      (3,2) &           59 &         0.291 &                35 &
$+40 + (11112) +2 (11122) -3 (11222) -4 (12222) +5 (22222)$ \\
      (4,1) &           21 &         0.291 &                 5 &   
$+16 - (11111) + (11112) + (11122) - (11222) - (12222) + (22222)$ \\
\bottomrule
\end{tabular}
\label{tab:fiveparty}
\caption{ For each hybrid model, the table shows the number of optimal symmetric
full-body correlation inequalities. Further a lower bound for the white-noise
robustness for the $|GHZ\rangle$ state is provided for the best inequality,
which is labeled by its number in the list in the Supplementary Material.}
\end{table*}

In the five-party scenario, we find nine inequalities for the fully local model,
$27$ inequalities for the $(2,1,1,1)$ model, $38$ inequalities for the $(2,2,1)$
model, $45$ inequalities for the $(3,1,1)$ model, $59$ inequalities for the
$(3,2)$ model and $21$ inequalities for the $(4,1)$ model. Among the Bell
inequalities in the $(3,2)$ model there are two Bell inequalities that are not
violated in quantum mechanics. For the $(4,1)$ model, there are nine such Bell
inequalities. All inequalities that are violated are maximally violated in
quantum mechanics by the GHZ state.

The Bell inequality for the fully local model
with the best white-noise robustness regarding the GHZ state is obtained for the
four-party Mermin inequality \cite{collins_bell-type_2002} 
\mm{
+4 - (11112) + (11222) - (22222) \ge 0. \label{eq-mermin5}
}
If the percentage of white noise in the state $\rho_5(p)$ is roughly smaller
than $74.3\%$, then the state exhibits nonlocality.

For the $(2,1,1,1)$ model, the most sensitive Bell inequality is the four-party
Svetlichny inequality \cite{collins_bell-type_2002}
\mm{
+8 &+ (11111) + (11112) - (11122)\notag\\ &- (22222) + (12222) + (11222) \ge 0.
\label{eq-svetlichny5}
}
It can detect a violation up to a white noise level up to roughly $64.5\%$. Note
that this is up to numerical precision the same threshold that we obtain for a
violation of the fully local model in the case of four parties.

We find numerically that the white-noise thresholds for two-local models,
three-local models and five-local models of the four-party scenario and the
five-party scenario coincide.
For the three-local models given by $h = (2,2,1)$ and $h = (3,1,1)$, we find that
the best inequality to detect the nonlocality of a noisy GHZ state is 
the generalized Mermin inequality \refeq{eq-mermin5}, however with a different bound
of eight instead of four in the fully local model. It can detect a violation up to
approximately $49.3\%$ of white noise.

The two-local models with $h = (3,2)$ and $h = (4,1)$ are violated up to
approximately $29.1\%$ of white noise. In the case of the $(4,1)$ model, the
violation is detected by the generalized Svetlichny inequality
\refeq{eq-svetlichny5} with an adapted bound of $16$. For the $(3,2)$ model, we
find that the most robust Bell inequality is
\mm{
+40 + (11112) &+2 (11122) -3 (11222) \notag\\&-4 (12222) +5 (22222) \ge 0. \label{eq-newine5}
}

\subsection{Family of hybrid Bell inequalities for many parties}
The inequalities \refeq{eq-newine4} and \refeq{eq-newine5} which 
are useful for ruling out a $(2,2)$ model or a $(3,2)$ model, respectively, can be extended to a family of Bell inequalities 
$(F_n)$ for an arbitrary number of $n$ parties. We define this
family as
\mm{
F_n = \sum_{\ell=1}^n (-1)^{1 + \lceil \frac{\ell}2 \rceil} \, \ell\, (\underbrace{1
\ldots 1}_{n-\ell}, \underbrace{2
\ldots 2}_{\ell}) \le n 2^{n-2}. \label{eq-newfam}
}
One can show that the bound can always be achieved in any $(k,m)$ model with $k, m>1$.
Moreover, for $m=2$, one can show analytically that the bound holds. 
Proofs for both statements can be found in Appendix \ref{app-bound}.
Moreover, we checked numerically for up to $20$ parties that the bound holds for $(k,m)$ models with $k>1, m>1$. 
We conjecture that the bound also tight for more parties.
For $(n-1, 1)$ models
the bound does not hold. A violation of an inequality in the family
\refeq{eq-newfam} therefore certifies that the nonlocality depth is at least
$(n-1)$.  

In the following, we discuss numerical findings concerning the Bell inequalities
$F_n$ for $n \le 8$ parties.
We find, that the quantum bound of the inequalities is 
\mm{
F_n = \sqrt 2 \, n 2^{n-2}
}
up to numerical precision, which is achieved by choosing $A_1 = B_1 = ... = \sigma_x$ and $A_2 = B_2 = ... = \sigma_z$, where $\sigma_x, \sigma_z$ are Pauli
matrices. The quantum states for which a maximal violation can be obtained using
these settings are listed in Appendix \ref{app-optimal}. We find numerically that they are
equivalent to the GHZ state up to local unitary transformations.

Further, we consider the white-noise robustness. For this, we consider the
states
\mm{
\rho_n(p) = (1-p) \ketbra{\text{GHZ}_n} + p \frac{\id_n}{2^n}.
}
We find that the inequalities are violated if $p\le 1 - 1/{\sqrt 2} \approx
29.3 \%$
up to numerical precision.

\section{Generalizations of a Bell inequality to more settings}
\subsection{What is a generalization of a Bell inequality?}
In this section we explain generalizations of Bell inequalities and present two
generalizations of the Svetlichny inequality.
Consider a Bell scenario $S_1$ and a Bell inequality $\mathbf b_1$ that is applicable to
the scenario $S_1$. Additionally, consider a second scenario, $S_2$, that is
larger than $S_1$ in the sense that it involves more parties, more
measurement settings per party or more outcomes per measurement setting than
$S_1$. Let $\mathbf b_2$ be a Bell inequality for $S_2$.
Consider the situation in which the parties perform a Bell test for the Bell
inequality $\mathbf b_2$. However, in the number of parties involved in the Bell test
and the number and kind of measurements they are allowed to perform, they
restrict themselves to the rules given by scenario $S_1$. If in this case the
Bell test effectively reduces to a Bell test of Bell inequality $\mathbf b_1$, then we
call $\mathbf b_2$ a generalization of $\mathbf b_1$.

In Ref.~\cite{bernards_generalizing_2020}, we introduced and discussed this concept in
the case of a generalization to more parties and describe a method how to find
generalizations of a given Bell inequality.
 The concept is best explained
by example. Consider three parties, $A, B, C$ that perform a test of Mermin's
inequality
\mm{
+ A_1 B_1 C_2+ A_1 B_2 C_1+ A_2 B_1 C_1- A_2 B_2 C_2 \le 2.
}
However, they restrict themselves to the resources of the CHSH scenario. This
means that one of the parties is no longer allowed to contribute to the Bell
test in a meaningful way. Let this party be $C$. If $C$ reports the measurement
outcome $+1$ in every round, then evaluating Mermins inequality effectively
means evaluating the inequality
\mm{
+ A_1 B_1 + A_1 B_2 + A_2 B_1 - A_2 B_2  \le 2,
}
which is the well-known CHSH inequality. Therefore, we call Mermin's inequality
a generalization of the CHSH inequality. To be precise, we call an inequality
$\mathbf b_2$ a generalization of an inequality $\mathbf b_1$ to more parties, if one can make a
choice of trivial measurements for the additional parties, such that $\mathbf b_2$
reduces to $\mathbf b_1$. By trivial measurement, we mean a measurement that yields one
measurement result deterministically. This has as a consequence, that each
$(n+1)$-party correlation can be computed from one $n$-party correlation, such
as
\mm{
\langle A_1 B_2 C_1 \rangle = \langle A_1 B_2 \rangle,\label{eq-extbeh0}
}
if $C_1$ is set to always yield outcome $+1$. 

We now turn to an example that illustrates the generalization to a scenario with
more settings. Consider the $(3,3;2,2)$ scenario, that is the bipartite
scenario, in which every party has three measurement settings and every setting
yields outcomes $\pm 1$. Besides the CHSH inequality, there is only one
non-trivial facet-defining inequality for this scenario, the I3322 inequality
\cite{froissart_constructive_1981, pitowsky_optimal_2001, collins_relevant_2004,
sliwa_symmetries_2003}. This inequality reads
\mm{
A_1 B_3 +
A_2 B_3 +
A_3 B_1 +
A_3 B_2 +
A_1  -
A_2  +
 B_1 -
 B_2 \notag \\-
A_1 B_1 +
A_1 B_2 +
A_2 B_1 -
A_2 B_2 
\le 4.
}
As was noted by Collins and Gisin \cite{collins_relevant_2004}, this inequality is
strictly stronger than the CHSH inequality with regard to its ability to detect
non-locality: If the measurements $A_3$ and $B_1$ are chosen trivially, so they
always yield the result $+1$, then the I3322 inequality reduces to a variant of
the CHSH inequality. Therefore the I3322 inequality is a generalization of the
CHSH inequality. This example also illustrates an important feature of
generalized Bell inequalities: A generalization of a Bell inequality always performs at least
as well as the Bell inequality itself in detecting nonlocality for a given
quantum state.

In the example of the I3322 inequality, $A$ and $B$ choose one of their measurements trivial to comply with the CHSH scenario.
In general, however, this is not the only way to achieve this. Alternatively, the parties may
have set two of their measurements equal up to a permutation of outcome labels. When looking for generalizations of a
Bell inequality to a scenario with more settings, one must therefore take this
possibility into account.

We shall now briefly discuss how to find generalizations of a Bell inequality
$\mathbf b_1$. Let $\mathbf b_1$ be a Bell inequality for scenario $S_1$ and one aims at finding
a Bell inequality $\mathbf b_2$ for a larger scenario $S_2$, such that $\mathbf b_2$ generalizes
$\mathbf b_1$. As a first step, we set the measurement settings that are present in
$S_2$ additionally to the ones present in $S_1$. As discussed earlier, the additional measurement
settings are either chosen trivial or equal to other measurement settings of the
same party up to outcome label permutations. For each behavior $\beta_1$ obtained in
scenario $S_1$, there is now a behavior $\beta_2$ for scenario $S_2$ that
corresponds uniquely to $\beta_1$, in the same fashion as in \refeq{eq-extbeh0}.
We call behavior $\beta_2$ the extended behavior of $\beta_1$.

The construction of extended behaviors allows us to express the property that
$\mathbf b_2$ is a generalization of $\mathbf b_1$ as a series of affine 
constraints: Let $\beta_1$ be a behavior that saturates $\mathbf b_1$. By 
this we mean that the behavior reaches the maximal classical value of one, which 
can also be written as a scalar product $\langle \beta_1, b_1 \rangle = 1$. Then, 
the extended behavior $\beta_2$ must saturate any generalization $\mathbf b_2$, 
that is $\langle \beta_2, b_2 \rangle = 1$. Since the set of all saturating 
behaviors of $\mathbf b_1$ defines the inequality $\mathbf b_1$ 
in a unique manner, the conditions formulated in this way are not only 
necessary but also sufficient. This means that if all extended behaviours 
obey $\langle \beta_2, b_2 \rangle = 1$, the $\mathbf b_2$ is a generalization 
of  $\mathbf b_1$.

So, facet defining inequalities of a polytope that obey a set of affine 
constraints can be found also using the CPT \cite{bernards_generalizing_2020, bernards_finding_2021}. Alternatively, if one is interested in finding the 
generalizing Bell inequality that is best suited to detect the nonlocality 
in a {\it given} behavior $r$, this is a linear program. One maximizes the 
expectation value over all inequalities $\mathbf b_2$ under two constraints. 
First, the inequality $\mathbf b_2$ has to obey $\langle \beta_2, b_2 \rangle = 1$
on all extened behaviours, as explained above. Second, $\mathbf b_2$ must be
a valid Bell-type inequality for all classical behaviours $\beta$ in the 
considered model of locality, i.e., $\langle \beta, b_2 \rangle$. More formally, 
this can be written as:
\mm{
&\max_{\mathbf b_2} \langle r, \mathbf b_2 \rangle \notag\\
\text{s.t.} \quad & \langle \beta_2, \mathbf b_2 \rangle = 1 \; \forall \text{ extended
saturating
behaviors } \beta_2, \notag \\
& \langle \beta, \mathbf b_2 \rangle \le 1 \; \forall \text{ extremal behaviors }\beta \text{ of the
model}.
\label{eq-linear_program}
}
Such a linear program can directly be solved using standard 
numerical techniques.

\subsection{Generalizations of the Svetlichny inequality to more settings}

Running the linear program \refeq{eq-linear_program} with random directions $r$, we find two generalizations of the Svetlichny inequality that are symmetric under
party permutations for the three-party scenario with three settings per party,
or $(3,3,3; 2,2,2)$ for short, 
\mm{
f_1 =& (100) - (111) + (211) + (221) - (222)\notag \\& + 2 (300) - (310) + (330) +
(331) \le 13 \\
f_2 =& -(122) + (123) + (133) - 3(222)\notag \\& - 2(223) + (233) \le 12.
}
The inequality $f_1$ reduces to the Svetlichny inequality, if one sets $A_3 = B_3
= C_3 = 1$. The second inequality, $f_2$, reduces to the Svetlichny inequality, if
one sets $A_3 = A_1, B_3 = B_1, C_3 = C_1$.

By construction, the inequalities $f_1, f_2$ are at least as sensitive to
nonlocality as the Svetlichny inequality. Moreover, since they have one more
setting, one might expect that there might be an advantage of $f_1$ and $f_2$
compared to the Svetlichny inequality. Unfortunately, sampling 540 random pure
three-qutrit states, we did not find a single example that shows an advantage.
Rather, we find that choosing the additional settings such that the inequalities
reduce to the Svetlichny inequality is always optimal. Accordingly, $f_1, f_2$
share the maximally violating state, the GHZ state, and their noise-robustness
with the Svetlichny inequality.

\section{Conclusion}

We presented the complete set of Bell inequalities to rule out 
various hybrid models for four- and five-body systems. These 
inequalities can be used to characterize the nonlocality depth
in these systems. Our analysis of GHZ states mixed with white 
noise suggests that the noise robustness of these states with
regard to a $k$-local model only depends on $k$. In contrast, 
the particular partition of parties that defines the $k$-local 
model seems to be irrelevant. For example, we did not find a 
difference between the $(3,2)$ model and the $(4,1)$ model in 
terms of noise-robustness.

Additionally to our analysis of four- and five-party scenarios, 
we presented a family of inequalities for an arbitrary number 
of parties $n$. The inequalities in this family are suitable 
to detect a nonlocality-depth of $n-1$. For $(k,m)$ models with 
$m>2$, we have a conjecture for the classical bound. Proving this 
bound or finding a counter-example remains an open problem.

Finally, we introduced the concept of a generalization of a Bell inequality to a scenario that involves more settings. We demonstrated
this concept by finding two inequalities that generalize the Svetlichny inequality. Unfortunately, these inequalities do not seem to have an advantage over the Svetlichny inequality. For future research, we believe 
it would be interesting to find generalizations of the Svetlichny 
inequality to more outcomes.

\acknowledgments
We thank Thomas Cope, Yeong-Cherng Liang, and Marc-Olivier Renou for 
ideas and discussions. 
This work was supported by the Deutsche Forschungsgemeinschaft (DFG, 
German Research Foundation, project numbers 447948357 and
440958198), the Sino-German Center for Research Promotion 
(Project M-0294), the ERC (Consolidator Grant 683107/TempoQ), 
and the House of Young Talents Siegen.

\appendix

\section{Description of Supplementary Material}
\label{app-supp}
The Supplementary Material available with the source code of this arxiv submission
consists of ten text files, one for each model
considered in this paper. For example, the file 'list221' lists all 38
inequalities we found for the $(2,2,1)$ model. The other files are named
analogously. For each inequality, we provide additional information such
as qubit and qutrit bound, no-signaling bound, and the bound provided by the
third level of the NPA hierarchy. For qubits, we additionally list the optimal
observables in the Pauli basis as well as the optimal quantum state in the
computational basis. 
Specifically the Supplementary Material consists of the following files:
\begin{description}
\item[A] list1111   contains details on the $(1,1,1,1)$ model.
\item[B] list211    contains details on the $(2,1,1)$ model.
\item[C] list22        contains details on the $(2,2)$ model.
\item[D] list31       contains details on the $(3,1)$ model.
\item[E] list11111  contains details on the $(1,1,1,1,1)$ model.
\item[F] list2111   contains details on the $(2,1,1,1)$ model.
\item[G] list221     contains details on the $(2,2,1)$ model.
\item[H] list311      contains details on the $(3,1,1)$ model.
\item[I] list32       contains details on the $(3,2)$ model.
\item[J] list41       contains details on the $(4,1)$ model.
\end{description}

\section{Estimating the noise-robustness interval}
 \label{app-interv}
To estimate the noise robustness, we calculate the maximal
violation of each inequality for states $\rho_n(p_i)$, where the values $p_i \in P$
are chosen equidistantly from the interval $[0,1]$. From this,
we obtain a critical interval $[p^c_0, p^c_1]$ that contains the noise
robustness $p^c$. Specifically, $p^c_0$ is defined as the largest possible value
in $P$, such that the maximal quantum violation of the Bell inequality exceeds
some threshold $t = 10^{-6}$. Similarly, $p^c_1$ is defined as the smallest
value in $P$, such that the Bell inequality is no longer violated. 
In a second step, we choose new, equidistant parameter values from
the critical interval and repeat the procedure. This algorithm is not very
efficient in the following sense. One can easily define an algorithm for which the
size of the critical interval decreases more quickly as a function of the number
of parameter values, for which the quantum violation of the Bell inequality is
computed. However, there is an advantage. The value computed for the quantum
violation is not guaranteed to be optimal. Calculating the quantum violation for
more parameters allows for a sanity check: The maximal quantum violation as a function
of the noise parameter $p$ is convex.

\section
{Classical bound for a family of Bell inequalities}
\label{app-bound}
In this section, we show that with any $(k,m)$ model, there exists a behavior
such that
\mm{
F_n = \sum_{\ell=1}^n (-1)^{1 + \lceil \frac{\ell}2 \rceil} \, \ell\, (\underbrace{1
\ldots 1}_{n-\ell}, \underbrace{2
\ldots 2}_{\ell}) = n 2^{n-2}.
}
Further, we show that for $m=2$, this bound is a valid upper bound of $F_n$.
First note that the symmetric correlation 
\mm{
(\underbrace{1
\ldots 1}_{n-\ell}, \underbrace{2
\ldots 2}_{\ell})}
  comprises $\binom{n}{\ell}$ terms, each of which takes
values $\pm 1$. Consequently,
\mm{
-\binom{n}{\ell} \le
(\underbrace{1
\ldots 1}_{n-\ell}, \underbrace{2
\ldots 2}_{\ell}) \le \binom{n}{\ell}.
}
Since the expression $F_n$ is linear in the symmetric correlation terms and its
maximum will therefore be achieved for 
\mm{
(\underbrace{1
\ldots 1}_{n-\ell}, \underbrace{2
\ldots 2}_{\ell}) = \pm \binom{n}{\ell}.
}
We can thus treat the symmetric correlation terms as binary variables. For
convenience, we define the variables
\mm{
\gamma^k_i = (\underbrace{1
\ldots 1}_{k-i}, \underbrace{2
\ldots 2}_{i}) \binom k i^{-1},
}
which are normalized such that they take values $\pm 1$. With this, we can
write $F_n$ as
\mm{
F_n = \sum_{\ell=1}^n (-1)^{1 + \lceil \frac{\ell}2 \rceil} \, \ell\,
\gamma^n_{\ell} \binom n{\ell}.
}
However, the variables $\gamma^n_{\ell}$ cannot be chosen independently, since
they have to respect the $(k,m)$ model under consideration. This condition is
met, if we consider behaviors that stem from a LHV model
between the first $k$ parties and the last $m$ parties. For this model, we have
\mm{
\gamma^n_{\ell} &= \gamma^k_i \gamma^m_j
\intertext{with}
i + j &= \ell.
}
With this, we can rewrite
\mm{
F_n = \sum_{\ell=1}^n (-1)^{1 + \lceil \frac{i+j}2 \rceil} \, (i+j)\,
\gamma^k_i \gamma^m_j \binom k{i} \binom mj.
}
Setting
\mm{
\gamma^k_i &= (-1)^{\lfloor \frac i2 \rfloor}\\
\gamma^m_j &= (-1)^{\lfloor \frac j2 \rfloor}
}
yields
\mm{
F_n &= \sum_{i=0}^k \sum_{j=0}^m (-1)^{(i+1)(j+1)} \, (i+j)\,
 \binom k{i} \binom mj\\
&=
n\, 2^{n-2}.
}
We now show that for $m = 2$, this value is a valid upper bound for $F_n$.
For convenience, we define the matrix $M$ with elements
\mm{
M_{ij} = (-1)^{(i+1)(j+1)} \, (i+j)\, \binom k{i} \binom 2j.
}
Note, that a different choice for $\gamma^k_i (\gamma^m_j)$ corresponds to flipping
the signs of the entries in the i-th row (j-th column) of $M$.
Hence, showing that there does not exist a subset of rows and columns, such that
multiplying these columns and rows with $-1$ yields a larger sum $\sum_{ij}
M_{ij}$ proves the claim.
Since $M$ only has three columns, we focus on the columns. For any choice of
rows and columns of $M$, either zero, one, two, or all columns of $M$ would be
affected. However, multiplying all columns and rows with $-1$ leaves $M$
invariant and therefore we only need to consider two cases: Either (1) non of the
columns is affected by the sign-flip operation or (2) exactly one column is affected
by the sign-flip operation. In case (1) one cannot reach a value higher than $n
2^{n-2}$ since all rows have a non-negative value. For the second case, note
that
\mm{
M_{i1} = |M_{i0}| + |M_{i2}|.
}
Hence, multiplying the column $j = 1$ with $-1$ cannot be compensated for any
choice of rows. Further, multiplying a column with $j\neq 1$ with $-1$ still
leaves all rows non-negative. Since the sum of the entries in the columns $j=0$
and $j=2$ vanishes, this means, that the value $F_n = n 2^{n-2}$ cannot be
exceeded in an $(n-2, 2)$ model.

\begin{widetext}
\section{Optimal states for Bell inequality family}
\label{app-optimal}
Below, we list the quantum states that lead to a maximal violation of the
respective Bell inequality from the $F_n$ family. For convenience, we define
\mm{
(X...Z) = X \otimes ... \otimes Z + \text{permutations},
}
where 'permutations' accounts for all party permutations of the first term and
no term is present twice in the sum, that is $(XXX)= X \otimes X \otimes X$. The
symbols $1,X,Y,Z$ are defined as
\mm{
1 = \frac 12 \id_2 , \;
X = \frac 12 \sigma_x, \;
Y = \frac 12 \sigma_y, \;
Z = \frac 12 \sigma_z.
}
The optimal states are the pure states
\begin{align}
\rho_3 = &(111)+ (1YY)-\frac 1{\sqrt{2}} (XXX)+\frac 1{\sqrt{2}} (XXZ)+\frac 1{\sqrt{2}}
(XZZ)-\frac 1{\sqrt{2}} (ZZZ) \\
 \rho_4 = &(1111)+ (11YY)-\frac 1{\sqrt{2}} (XXXX)+\frac 1{\sqrt{2}}
(XXXZ)+\frac
1{\sqrt{2}} (XXZZ)-\frac 1{\sqrt{2}} (XZZZ)+ (YYYY)-\frac 1{\sqrt{2}} (ZZZZ)\\
 \rho_5 =& (11111)+ (111YY)+ (1YYYY)-\frac 1{\sqrt{2}} (XXXXX)+\frac 1{\sqrt{2}}
(XXXXZ)+\frac 1{\sqrt{2}} (XXXZZ)-\frac 1{\sqrt{2}} (XXZZZ)\notag\\ &-\frac 1{\sqrt{2}}
(XZZZZ)+\frac 1{\sqrt{2}} (ZZZZZ) \\
 \rho_6 =& (111111)+ (1111YY)+ (11YYYY)-\frac 1{\sqrt{2}} (XXXXXX)+\frac 1{\sqrt{2}}
(XXXXXZ)+\frac 1{\sqrt{2}} (XXXXZZ)\notag\\ &-\frac 1{\sqrt{2}} (XXXZZZ)-\frac 1{\sqrt{2}}
(XXZZZZ)+\frac 1{\sqrt{2}} (XZZZZZ)+ (YYYYYY)+\frac 1{\sqrt{2}} (ZZZZZZ)\\
 \rho_7 =& (1111111)+ (11111YY)+ (111YYYY)+ (1YYYYYY)-\frac 1{\sqrt{2}} (XXXXXXX)+\frac
1{\sqrt{2}} (XXXXXXZ)\notag\\ &+\frac 1{\sqrt{2}} (XXXXXZZ)-\frac 1{\sqrt{2}}
(XXXXZZZ)-\frac 1{\sqrt{2}} (XXXZZZZ)+\frac 1{\sqrt{2}} (XXZZZZZ)+\frac
1{\sqrt{2}} (XZZZZZZ)\notag\\ &-\frac 1{\sqrt{2}} (ZZZZZZZ)\\
 \rho_8 =& (11111111)+ (111111YY)+ (1111YYYY)+ (11YYYYYY)-\frac 1{\sqrt{2}}
(XXXXXXXX)\notag\\ &+\frac 1{\sqrt{2}} (XXXXXXXZ)+\frac 1{\sqrt{2}} (XXXXXXZZ)-\frac
1{\sqrt{2}} (XXXXXZZZ)-\frac 1{\sqrt{2}} (XXXXZZZZ)\notag\\ &+\frac 1{\sqrt{2}}
(XXXZZZZZ)+\frac 1{\sqrt{2}} (XXZZZZZZ)-\frac 1{\sqrt{2}} (XZZZZZZZ)+
(YYYYYYYY)-\frac 1{\sqrt{2}} (ZZZZZZZZ)
\end{align}
This can be generalized to
\mm{
\rho_n =& \sum_i (1^{n-2i} Y^{2i}) + \frac 1{\sqrt 2} \sum_l (-1)^l [ (X^{2l} Z^{n-2l}) -
(X^{2l + 1} Z^{n-2l-1})],
\intertext{which is equivalent to}
\equiv& \sum_i (1^{n-2i} Z^{2i}) + \frac 1{\sqrt 2} \sum_l (-1)^l [ ( X^{n-2l}Y^{2l}) -
(X^{n-2l-1} Y^{2l + 1} )]
\intertext{under the local unitary transformation}
U =& \frac 1{\sqrt 2} \begin{pmatrix} 1& -i \\ 1 & i \end{pmatrix}.
\intertext{For comparison, the standard GHZ state written in the z-basis reads}
\rho_{ghz,n} = &\sum_i (1^{n-2i} Z^{2i}) + \sum_l (-1)^l  ( X^{n-2l}Y^{2l}).
}
Numerically, we find that the optimal states are equivalent under local unitary
transformations to GHZ states.

\end{widetext}

\bibliography{bibliothek}
\bibliographystyle{apsrev4-2}

\end{document}